\documentclass [12pt]{article}
\usepackage{lscape}                                           %
\usepackage[centertags]{amsmath}
\usepackage{amsfonts} \usepackage{amssymb}\usepackage{eufrak}
\usepackage{graphicx}
\usepackage{a4}
\usepackage{cite}
\usepackage{lscape}
\begin{document}
\begin{titlepage}
\rightline{}
\rightline{}
\vskip 3.0cm

\centerline{\LARGE \bf Magnetized Branes and the Six-Torus\footnotetext[1]{
Proceedings of the Corfu Summer Institute 2012 ``School and Workshops on Elementary Particle Physics and Gravity'', September 8-27, 2012, Corfu, Greece.}}
\vskip .5cm

\vskip 1.0cm
\centerline{\bf L. De Angelis$^a$, R. Marotta$^b$, F. Pezzella$^b$ and  R. Troise$^a$\footnote[2]{XXIII  Ciclo di Dottorato in Fisica Fondamentale e Applicata.}}
\vskip .6cm \centerline{\sl  $^a$ Dipartimento di
Scienze Fisiche, Universit\`a degli Studi ``Federico II'' di Napoli}
%\centerline{\sl XXIII  Ciclo di Dottorato in Fisica Fondamentale e Applicata}
\centerline{\sl Complesso Universitario Monte
S. Angelo  ed. 6, via Cintia,  80126 Napoli, Italy}
\vskip .4cm
\centerline{\sl $^b$ Istituto Nazionale di Fisica Nucleare, Sezione di Napoli}
\centerline{ \sl Complesso Universitario Monte
S. Angelo ed. 6, via Cintia,  80126 Napoli, Italy}
\vskip 0.4cm

\begin{abstract}
In the framework of Type IIB String  Theory compactified on a general six-torus $T^6$ with arbitrary complex structure,  Yukawa couplings are determined for the chiral matter described by open strings ending on D9-branes having different oblique magnetization.
\end{abstract}

\end{titlepage}

\newpage
\tableofcontents
\vskip 1cm

\section{Introduction}
Many new
ideas and tools have been  introduced in the last years in order to connect the ten-dimensional  Superstring Theory to  the four-dimensional Standard Model. This is the goal of the so-called {\em String Phenomenology} which in fact studies how  the (beyond)  Standard Model physics could be obtained as a low-energy limit of String Theory \cite{IU,Review1,Review2}.

 In orientifolds of Type II String Theory, the main ingredients that String Phenomenology uses for achieving its aim are Dp-branes together with the compactification of the extra dimensions and the derivation of chiral spinors. In particular, the gauge groups of the Standard Model are localized in the world-volume of suitable configurations of Dp-branes and the chiral matter can be introduced by ``dressing''  the compact directions with magnetic fields \cite{magnetized1,magnetized3,BachasPorrati,0512067,0709.4149,also6}. The open strings ending on different piles of branes
with different magnetizations are named {\em dy-charged} or {\em twisted} strings and they exactly  describe the chiral matter of the low-energy theory.

In order to promote magnetized branes in a compact space as vacua interesting for the String Phenomenology,  one needs to have information about the low-energy effective actions living in the world-volume of such configurations of branes \cite{0508043,also4, also5}. These actions are model dependent  and a particular interest, especially after the Higgs discovery, is focused on the determination  of the Yukawa couplings
and their dependence on the details of the compact manifold, i.e.  moduli and geometry of the extra dimensions. Yukawa couplings in Type II brane-world  scenario arise from an overlap integral of wave-functions of the three participant fields in the extra dimensions. The
wave-functions, depending on the Bose-Fermi statistics of the fields, are solutions of the internal Laplace-Beltrami or Dirac equation with suitable boundary conditions dictated by the compact geometry and by the presence of the magnetic fields. In the {\em bottom-up} approach one usually neglects the global aspects of the compactification  and solves these equations locally. However, for the simplest compact manifold, the factorized torus,  the boundary conditions imposed by the magnetized torus geometry have exactly determined  the holomorphic part of the Yukawa couplings that turns out to be
%that the  holomorphic part of the  Yukawa couplings is
proportional to  the Jacobi $\theta$-function \cite{0404229,0701292,0807.0789,0810.5509,0906.3033}.
The global properties  of the compact manifold  are then important to fix the complete structure of the effective actions and it results to be interesting to compute such couplings in the case of non factorized geometries as the one of the torus $T^6$ with arbitrary complex structure.

In this talk - which is based on the paper of ref. \cite{1206.3401} - these couplings are studied in a configuration of  $M$ D9-branes in the  background $\mathbb{R}^{1,3}\times T^6$. In the same spirit as the one of ref.
\cite{0404229} (see also \cite{0810.5509,1101.0120}), constant magnetic fields are turned on, along the compact directions,  in the abelian sector of the $U(M)$ gauge group
defined on the world-volume of the $M$  branes. Depending on the choice of such constant
fields, the single stack of branes is now separated in  different piles
of magnetized branes.
The ten-dimensional ${\cal N}=1$ super Yang-Mills theory living in the world-volume of a
stack of D9-branes is dimensionally reduced to four dimensions by
expanding the ten-dimensional bosonic or fermionic fields in a basis of
eigenfunctions of the internal Laplace or Dirac operator.
The eigenfunctions of these operators  have to be
invariant, up to gauge transformations, when translated along the one-cycles of the torus. They are easily determined  in the complex frame
where both the metric and the difference $ F^{ab}=F^a-F^b$ of the magnetic fields on the
two piles $a$ and $b$ of branes between which  the strings are stretched,  are diagonal matrices in their off-diagonal boxes. In this frame, supersymmetry has been also partially imposed by requiring the field
$F^{ab}$ to be  a $(1,1)$-form in the coordinate system defining the complex torus.
The wave-functions of twisted  open strings turn out to be proportional to  the Riemann Theta function only when the background gauge field, in the original
system of coordinates defining the torus, is a matrix with null diagonal blocks.
They depend on the first Chern class $I_{ab}$ associated with the difference of the gauge fields on the $a$ and $b$ branes  and on  a  generalized complex structure that is a matrix whose  entries are  related
to the original complex structure of the torus  or to  its complex
conjugate, depending on the signs of the eigenvalues of the non-vanishing blocks of the gauge field $F^{ab}$.

The Yukawa couplings are obtained by evaluating an overlap integral over three of such functions. The integral  is  computed  after using  an identity between the product of two Riemann $\theta$-functions. The identity has been  derived in refs. \cite{1206.3401, 0904.0910} by extending the analysis  given in ref. \cite{MumfordI} and here revised. The resulting expression is compatible with the known results obtained under different assumptions \cite{0709.1805, 0904.0910};  the non trivial and holomorphic part of these couplings, the Riemann  $\theta$-function,  is again determined by the boundary conditions due to the geometry of the magnetized torus. Here holomorphicity  means,  as in the factorized case, that the $\theta$-function can never depend on a variable and its complex conjugate, although the argument of such a  function can be either holomorphic or antiholomorphic along different directions. These properties are related to the signs of the first Chern classes evaluated along the corresponding compactified directions of the torus.

The paper is organized as follows.

In section  \ref{2}, generalities about dimensional reduction and magnetic fluxes are given.
In section \ref{3}, the bosonic and fermionic wave-functions for the lowest states are derived together with the mass spectrum of the Kaluza-Klein  states. In section \ref{4}, the Yukawa couplings for a general magnetized six-torus $T^6$ are computed.  Finally, in the appendix details about the proof of an identity involving the product of two wave-functions are given.

\section{Open Fluxes and Torus Geometry}
\label{2}
A configuration made of a stack of $M$ D9-branes in the compact background $\mathbb{R}^{1,3}\times T^6$ is going to be studied in this paper. Branes backreaction on the space-time  geometry is neglected and the analysis is focused on the open string degrees of freedom. Their interaction with the closed string degrees of freedom is described by the supersymmetric  DBI and by the Chern-Simons actions. In the following, attention will be drawn to the low-energy limit of the DBI action which, for this particular brane configuration, is  the ten-dimensional ${\cal N}=1$ super Yang-Mills with gauge group $U(M)$:
\begin{eqnarray}
S = \frac{1}{g^2} \int d^{10} X^{ \hat{N}} \,\, {\mbox Tr}
\bigg(- \frac{ 1}{ 4} {\cal F}_{\hat{M}\hat{N}} {\cal F}^{\hat{M}\hat{N}}  +
\frac{i}{  2}\bar{\lambda} \Gamma^{\hat{M}} {D}_{\hat{M}} \lambda \bigg)
\label{10dimla}
\end{eqnarray}
where $\hat{M}, \hat{N} = 0,\dots,9$, \, $g^2 =  4 \pi {\rm e}^{\phi_{10}} (2 \pi \sqrt{\alpha'})^6$ and
\begin{eqnarray*}
{\cal F}_{\hat{M}\hat{N}} = \nabla_{\hat{M}} A_{\hat{N}} - \nabla_{\hat{N}} A_{\hat{M}}  - i[A_{\hat{M}}, A_{\hat{N}}]~~;~~
{D}_{\hat{M}} \lambda = \nabla_{\hat{M}} \lambda - i [A_{\hat{M}},\lambda]
\end{eqnarray*}
with $ \lambda$ being  a ten-dimensional Weyl-Majorana spinor.
Chiral matter is introduced by turning on magnetic fields with constant field strength along the compact directions of the world-volume of $N_a$ branes, with $\sum_{a=1}^nN_a=M$.   The integer $n$ labels the branes having different  magnetization.  The original gauge group is then broken into the product $U(M)\simeq \prod_{a=1}^nU(N_a)$ and this breaking can be  used to engineer  the gauge groups of the Standard Model. The chiral matter is given by the twisted open strings charged with respect two of these groups and transforms in the bifundamental representation of the gauge group $U(N_a) \times U(N_b)$. In the following, the complete breaking $U(M)\simeq U(1)^M$ is considered, but  the extension to other gauge configurations  is straightforward.
The breaking  is realized by first separating the generators $U_a$ of the Cartan subalgebra from the ones out of it,  $e_{ab}$, in the definitions of the gauge field and of the gaugino:
\begin{eqnarray*}
A_{\hat{M}} = B_{\hat{M}} + W_{\hat{M}} = B_{\hat{M}}^a U_a + W_{\hat{M}}^{ab} e_{ab}~~;~~
\lambda  =  \chi + \Psi = \chi^{a}U_a + \Psi^{ab}e_{ab}
\end{eqnarray*}
and then expanding the Lagrangian around the background fields which are present only along the compact directions in $T^{6}$ of the branes:
\begin{eqnarray}
B_M^{a}(x^{\mu} , X^{N} ) & = & \langle B_M^{a} \rangle (X^N)+
 \delta B_M^{a}(x^{\mu} , X^N) \nonumber \\
W_M^{ab}(x^{\mu} , X^{N}) & = & 0 + \Phi_M^{ab}(x^{\mu}, X^N) \,\, .
\label{expa3}
\end{eqnarray}
Here $\mu = 0, \dots, 3$ and $M, N=1, \dots, 6$. The fields $B^{a}_M$ and $\Phi_M^{ab}$ are, respectively, adjoint and chiral scalars, from the  point of view of the four-dimensional Lorentz group.  The background fields $\langle B^{a}_{M}\rangle $ are taken with a constant field strength corresponding to the background constant magnetic fields along the compact dimensions. In particular, the gauge
\begin{eqnarray}
\langle B^{a}_{M} \rangle (X^{N}) = - \frac{1}{2} F_{MN}^{a} X^{N}\,\,  \nonumber
\end{eqnarray}
is chosen.

The four-dimensional effective action is obtained by compactifying the extra dimensions
on the torus $T^6$,  defined by imposing  the identification
\begin{eqnarray}
x^m\equiv x^m+2\pi R_1^{(m)}m_1^m~~;~~y^m\equiv y^m+ 2\pi\,R_2^{(m)}m_2^m~~~~m_1^m,\,m_2^m\in \mathbb{Z}\nonumber
\end{eqnarray}
on the space-time coordinates $(x^m,\,y^m)\equiv (X^{2m-1},\,X^{2m})$ ($m=1,2,3$), being $R_1^{(m)}$ and $R_2^{(m)}$ the radii  of the torus along the direction $m$. In the following, in order to compare the string with the field theory results, it is convenient to use the following rescaling:
\begin{eqnarray}
(x^{m}, y^{m}) \rightarrow \left( x^{m} \frac{\,\,\,\,\, R_{1}^{(m)}}{R}, \,\, y^{m} \frac{\,\,\,\,\, R_{2}^{(m)}}{R} \right) ,\nonumber
\end{eqnarray}
with $R$ being  an arbitrary  dimensionful parameter, and then to define the torus geometry  through the identification:
\begin{eqnarray}
x^m\equiv x^m+2\pi R\,m_1^m~~;~~y^m\equiv y^m+2\pi\, R\,m_2^m\,\, .\nonumber
\end{eqnarray}
The description of the  torus as  a complex manifold is based on the  introduction of the coordinates:
\begin{eqnarray}
w^m=\frac{x^m+U^m_{~n}\,y^n}{2\pi R}\nonumber
\end{eqnarray}
 together with their complex conjugate variables. Here, $U$ is a complex matrix parametrizing the complex structure of the manifold.  The lattice identification is given by
\begin{eqnarray}
w^m \equiv w^m+m_1^m+U_{~n}^m\,m_2^n \,\, .\nonumber
\end{eqnarray}
The twisted sectors of the theory, as previously discussed, are present because a background magnetic field  $F^{ab}=F^a-F^b$  acts on the world-volume of two piles of branes $a$ and $b$. In the system of complex coordinates it takes the following form \cite{1206.3401}:
\begin{equation}
F^{ab}=-\frac{(2\pi R)^2}{8}  F^{{(\cal{WW}}) ab}_{MN} d{{\cal{W}}^M} \wedge   d{{\cal{W}}^N} \nonumber
\end{equation}
with $( {\cal W}^{1}, \dots, {\cal W}^{6}) \equiv (w^1, \dots, w^3, \bar{w}^1, \dots \bar{w}^3)$ and
\begin{eqnarray}
&&F^{(ww)ab}= ({\rm Im}U^{-1})^t \left[\bar{U}^t \,F^{(xx)ab} \bar{U} -\bar{U}^t F^{(xy)ab} +{F^{(xy)ab\,t}}\bar{U} +F^{(yy)ab}\right] {\rm Im}U^{-1} \nonumber\\
&&{ F}^{(w\bar{w})ab}= ({\rm Im}U^{-1})^t \left[-\bar{U}^t F^{(xx)ab} U+\bar{U}^t F^{(xy)ab} -{F^{(xy)ab\,t}}U-F^{(yy)ab}\right] {\rm Im}U^{-1} \nonumber
\end{eqnarray}
while ${F}^{(\bar{w}w)ab}= [F^{(w\bar{w})ab}]^*$ and ${F}^{(ww)ab}= [F^{(\bar{w}\bar{w})ab}]^*$; furthermore the index $t$ denotes the transposed of the matrices it refers to.  Supersymmetric configurations require the gauge field to be a $(1,1)$-form. Imposing such constraint necessarily makes
$iF^{(w \bar{w})}$  an Hermitian matrix \cite{0904.0910} which is diagonalized by an unitary matrix $\bar{C}_{ab}^{-1}$:
\begin{eqnarray}
(C_{ab}^{-1})^m_{~r}\,\, F_{m \bar{n}}^{(w, \bar{w})ab} \,\, (\bar{C}_{ab}^{-1})^{{\bar{n}}}_{~\bar{s}} = \frac{2}{i} \frac{\lambda_r^{ab}}{(2\pi R)^2} \delta_{r\bar{s}} \,\, . \nonumber
\end{eqnarray}
Here, $r,s=1,\dots 3$  and,  since $\bar{C}_{ab}^{-1}$ is an unitary matrix, one has $(C^{-1}_{ab})^m_{~r} h_{m\bar{n}} (\bar{C}_{ab}^{-1})^{{\bar{n}}}_{~\bar{s}}=\delta_{rs}$ where $h_{m\bar{n}}$ refers to the metric of the complex torus which  can also be written in terms of the holomorphic and anti-holomorphic vielbeins: $h_{m\bar{n}}=e^r_{~\bar{m}} \delta_{r\bar{s}}\bar{e}^{\bar s}_{~\bar{n}}$. Complex coordinates having a trivial metric can now be introduced by defining:
 $w^r=e^r_{~m}w^m$ together with their complex conjugate.

A new  system of complex coordinates $({\cal Z}_{ab}^1,\dots, {\cal Z}_{ab}^{6})~=~({z}_{ab}^{1}, \dots, {z}_{ab}^{3}, \,{\bar z}_{ab}^1,\dots,  {\bar z}_{ab}^3 )$, defined by
\begin{eqnarray}
w^m= \left(C^{-1}_{ab}\right)^m_{~r}{z}^r_{ab}~~;~~\bar{w}^{m}= \left(\bar{C}^{-1}_{ab}\right)^{{\bar{s}}}_{~\bar{r}}\bar{ z}^{r}_{ab} \,\, , \nonumber
\end{eqnarray}
is naturally introduced by the diagonalization.
 In this frame, both the metric and the field strengths of the gauge field are diagonal matrices in the non-vanishing blocks, being the metric equal to:
\begin{eqnarray}
\frac{ds^2}{(2\pi R)^2} =\frac{1}{2}\, d{\cal Z}^I_{ab}\, {\cal G}_{IJ} \,  d{\cal Z}_{ab}^J~~;~~
{\cal G}=\left( \begin{array}{cc}
0&\mathbb{I}\\
\mathbb{I}&0\end{array}\right)\nonumber
\end{eqnarray}
while the  magnetic field strength is
\begin{eqnarray}
F^{ab}
&=& \frac{1}{2} F^{({\cal Z} {\cal Z}) ab}_{IJ} d{\cal Z}_{ab}^I\wedge d{\cal Z}_{ab}^J~~;~~F_{IJ}^{({\cal Z} {\cal Z}) ab}= \frac{i}{2}\left(\begin{array}{cc}
0& {\cal{I}}_{\lambda}^{ab}\\
-{\cal{I}}_{\lambda}^{ab}&0\end{array}\right)\label{89}
\end{eqnarray}
with
\begin{eqnarray}
{\cal{I}}_{\lambda}^{ab}={\rm diag} \left( \lambda_1^{ab}  \dots  \lambda_{d}^{ab} \right)\nonumber
\end{eqnarray}
 and $I,J=1, \dots, 6$ denote the flat indices.

\section{The Wave-Functions}
\label{3}
The quadratic terms in the scalar fields  of the four-dimensional action, derived in detail in
ref. \cite{0404229}, see also \cite{0807.0789,0810.5509}, are  obtained by starting from  eq. (\ref{10dimla}) and expanding  the fields,  defined in the second line of eq. (\ref{expa3}),  in a basis of eigenfunctions of the internal Laplace-Beltrami operator:
\begin{eqnarray}
- \tilde{D}_N\tilde{D}^N \phi_{\cal M}^{ab} (X^N)=m_{{\cal M}}^2\phi^{ab}_{{\cal M}}(X^N) ~~;~~ \Phi_M^{ab}= \sum_{\cal M} \varphi^{ab}_{M, \,{\cal M}}(x^\mu) \otimes \phi^{ab}_{\cal M}(X^N)   \nonumber
\end{eqnarray}
with suitable boundary conditions determined by the torus geometry. Here, the
 covariant derivative depends only on the constant background gauge fields
\begin{eqnarray}
\tilde{D}_N\phi^{ab}_{\cal M}=\partial_N\phi^{ab}_{\cal M}-i  (\langle B_{N}^{a} \rangle-\langle B_{N}^b \rangle)\phi^{ab}_{\cal M}\,\, .  \nonumber
\end{eqnarray}
The mass spectrum of the Kaluza-Klein  states is easily determined in the system of the ${\cal Z}$-coordinates. In such a frame the mass operator is \cite{1206.3401}:
\[
\left[ M^{2}_{\cal M} \right]^{ab}  \equiv \mbox{ diag}~(\tilde{m}_{\cal M}^{2ab }~ \mathbb{I}~- ~2~ {\cal I}_{\lambda}^{ab}, \tilde{m}_{\cal M}^{2ab} \mathbb{I} +~2~ {\cal I}_{\lambda}^{ab} )
\]
with $\tilde{m}_{\cal M}^{ab}= 2\pi R \,\, m_{\cal M}^{ab}$ while, due to the block diagonal expression of the background gauge field,
the commutation relations $[\tilde{D}_{I}^{{(\cal Z)}},\,\tilde{D}_{J}^{{(\cal Z})}]=-i\,F_{IJ}$  of the covariant derivatives reduce to the algebra of  decoupled creation and annihilation bosonic operators. This identification depends on the signs of the eigenvalues $\lambda_r$, being for positive $\lambda_r$ \footnote{In this analyis the $a,b$ labels are omitted when possible.}:
\begin{eqnarray}
a_r^\dagger = \sqrt{\frac{2}{|\lambda_r|}}i\,\tilde{D}_r^{({\cal Z})}~~;~~a_r= \sqrt{\frac{2}{|\lambda_r|}} i \tilde{D}_{r +d}^{({\cal Z})} \label{osc0}
\end{eqnarray}
with the role of the creation and annihilation operators exchanged for negative $\lambda_r$. In both cases one has $[a_r,\,a^\dagger_r]=1$ and the Laplace equation becomes
\begin{eqnarray}
\sum_{r=1}^3 |\lambda_r| (2N_{r} + 1) \phi_{\cal M} =\tilde{m}^2_{\cal{M}} \phi_{\cal M}~~;~~N_r=a^\dagger_r\, a_r \,\, .   \nonumber
\end{eqnarray}
The eigenvalues of the mass operator result to be:
\begin{equation}
M^{2}_{\pm ; s} = \sum_{r=1}^3  |\lambda_r| (2\, N_r+1) \mp 2 \lambda_{s}  \label{m2}  \,\, .\nonumber
\end{equation}
The lightest state is massless if the ${\cal N}=1$ susy condition $|\lambda_r|+|\lambda_s|=|\lambda_t|$ ($r\neq s\neq t$) is imposed. Then, by applying  creation operators on the massless state, two towers of Kaluza-Klein states are generated.  Their  spectrum, when the ${\cal N}=1$  susy condition is imposed, is contained in the expression:
\begin{eqnarray}
M^2_k=  2 \sum_{r=1}^{3} |\lambda_r|(N_r+k)~~;~~k=0,\,1\,\, .  \label{bms}
\end{eqnarray}
The eigenfunctions relative to the ground state are obtained by solving the first order differential equation
\begin{eqnarray}
a_r\phi_0=0 ~\forall r~~ \Leftrightarrow ~~\left(\frac{\partial}{\partial \bar{\mbox{z}}^r}+\frac{1}{4} |\lambda_r|\,\mbox{z}^r\right)\phi_0=0\nonumber
\end{eqnarray}
where the complex coordinates $(\mbox{z}^1,\dots,\mbox{z}^3,\,\bar{\mbox z}^1\dots,\bar{\mbox z}^3)$, with
\begin{eqnarray}
\mbox{z}^r=z^r(\frac{1+{\rm sign}(\lambda_r)}{2})+\bar{z}^r(\frac{1-{\rm sign}(\lambda_r)}{2})=
(C^{(\lambda)})^r_{~m}\left(\frac{x^m
+\Omega^m_{~n}y^n}{2\pi R}\right)  \label{Cr} \,\, ,
 \end{eqnarray}
 have been introduced in order to take into account that the identification between the covariant derivatives and the  creations or annihilations operators depends on the signs of the eigenvalues $\lambda_r$.
 In eq. (\ref{Cr})
\begin{eqnarray}
{C^{(\lambda)}}^r&=&\left(\frac{1+{\rm sign} (\lambda_r)}{2}\right)
C^r+ \left(\frac{1-{\rm sign} (\lambda_r)}{2}\right)\bar{C}^r=\bar{C}^{(-\lambda)\,r}\nonumber\\
{\tilde C}^{(\lambda)\,r}&=&\left(\frac{1+{\rm sign} (\lambda_r)}{2}\right)
C^r  U+ \left(\frac{1-{\rm sign} (\lambda_r)}{2}\right)\bar{C}^r \bar{U}=\bar{\tilde C}^{(-\lambda)\,r}\label{defcl}
\end{eqnarray}
and $\Omega= (C^{(\lambda)})^{-1}  \tilde{C}^{(\lambda)}$ is named the {\em generalized complex structure} because, when all the $\lambda_r$s have  the same sign,  it coincides with the complex structure of the torus or its complex conjugate.

The wave-function of the ground state is:
\begin{eqnarray}
\phi_0=e^{-\frac{1}{4} \vec{\bar{\mbox  z}}^t  |{\cal I}_{\lambda}|\vec{ \mbox z} +\frac{1}{4}\vec{\mbox z}^t {C^{(\lambda)}}^{-t}\bar{C}^{(\lambda)\,t} |{\cal I}_\lambda| \vec{ \mbox z}}
\theta(\vec{\mbox z})\label{ansatz1}
\end{eqnarray}
where $(Z^1,\,\dots,Z^6)\equiv (\mbox{z}^1,\dots,\mbox{z}^3,\,\bar{\mbox z}^1\dots,\bar{\mbox z}^3)$  and with $\theta(\vec{\mbox{z}})$ being an holomorphic function of the coordinates which is determined by  the boundary conditions. It is interesting to notice that the wave-function (\ref{ansatz1}), when rewritten in the original system of coordinates ${\cal Z}^I=(z^r,\bar{z}^r)$, may depend on both  the holomorphic and anti-holomorphic variables. However, it never simultaneously depends on a variable and its complex conjugate, i.e. on $z^r$ and $\bar{z}^r$ (same $r$) and,  therefore $\theta$ is an holomorphic function of the  complex coordinates.

Boundary conditions  are dictated by the transformation properties  of the scalar fields under the torus translations \cite{THooft, 0306006}.
The behavior  of the vector potential $A^{ ( \mbox{z} ) }_{r}$ under the lattice translations
\begin{eqnarray}
A^{ (\mbox{z} )}_{r} (\bar{\mbox{z}}+ \bar{C}^{(\lambda)} \eta_{(s)})\equiv A^{ ({\mbox{z})}}_{r}( \bar{\mbox{z}})+\partial_{r}\chi^{(1)}_{(s)}~~;~~A^{ (\mbox{z} )}_{r}( \bar{\mbox{z}}+ \bar{{C}}^{(\lambda)}\, \Omega \,\eta_{(s)})\equiv A^{ (\mbox{z} )}_{r}( \bar{\mbox{z}})+\partial_{r}\chi^{(2)}_{(s)}  \nonumber
\end{eqnarray}
defines the corresponding  gauge transformations
\begin{eqnarray}
\chi_{(s)}^{(1)}&=& -\frac{i}{4} {\mbox{z}}^r|\lambda_r|  (\bar{C}^{(\lambda)})^r_{~n}\eta^n_{(s)}+\frac{i}{4} \bar{\mbox{z}}^r|\lambda_r|  ({C}^{(\lambda)})^r_{~n}\eta^n_{(s)}\nonumber\\
\chi_{(s)}^{(2)}&=& -\frac{i}{4} {\mbox{z}}^r|\lambda_r|  ({\bar{C}}^{(\lambda)})^r_{~m}\,\bar{\Omega}^n_{~m}\,\eta^n_{(s)}+\frac{i}{4} \bar{\mbox{z}}^r|\lambda_r|  ({ C}^{(\lambda)})^r_{~m}\,\Omega^m_{~n}\eta^n_{(s)} \,\, \nonumber
\end{eqnarray}
with $\eta^t_{(s)}=(\overbrace{0,\dots,0, 1}^{s \,\,\, times},0,\dots)$.
The holomorphic function appearing in the definition of the ground state is determined by imposing the identifications
\begin{eqnarray}
\phi_0(\vec{\mbox z}+ {C^{(\lambda)}} \eta_{(s)},\,\vec{\bar{ \mbox z}}+\bar{C}^{(\lambda)} \eta_{(s)})=e^{i\chi_{(s)}^{(1)}}
\phi_0(\vec{\mbox z},\,\vec{\bar{ \mbox z}})\nonumber\\
\phi_0(\vec{ \mbox z}+ {C}^{(\lambda)}\, {\Omega}\,  \eta_{(s)},\,\vec{\bar{ \mbox z}}+\bar{ C}^{(\lambda)}\,\bar{\Omega}\,  \eta_{(s)})=e^{i\chi_{(s)}^{(2)}}
\phi_0(\vec{\mbox  z},\,\vec{\bar{\mbox z}}) .  \,\, \label{bctorus}
\end{eqnarray}
The full wave-function of the ground state, in the real coordinates system and in the case $F^{(xx)}=F^{(yy)}=0$, is
\begin{eqnarray}
\phi_0\equiv\phi_{\Omega_{ab};\,\vec{j}}^{ab}(x^m,\,y^m)= {\cal N}_{ab} \,e^{i\vec{y}^t\,\frac{I_{ab}}{(2\pi R)^2} \, \vec{x}+i \vec{y}^t \Omega_{ab}^t\,\frac{  I_{ab}^t }{(2\pi R)^2}\vec{y}}
\sum_{\vec{n}\in \mathbb{Z}^{d}}
e^{ i \pi (\vec{n}+\vec{j})^t  I_{ab} \Omega_{ab}  (\vec{n}+\vec{j})+  2 i \pi (\vec{n}+\vec{j})^t  I_{ab} (\frac{\vec{x}+\Omega_{ab} \, \vec{y}}{2\pi R})}\label{83}
\end{eqnarray}
where  the  overall constant ${\cal N}_{ab}=\sqrt{2g}V_{T^{2d}}^{-1/2}\left[ {\rm det}( I_{ab}{\rm Im}\Omega_{ab})\right]^{1/4} $  is fixed by requiring canonical normalization for the kinetic terms of the scalars. Here, $V_{T^{2d}}$ is  the torus volume and $I= 2\pi \,R^2 {F^{(xy)t}}$.

The wave-function (\ref{83}) can be easily compared with the corresponding expression given in  ref. \cite{0904.0910} for  the torus $T^4$. The two expressions coincide if the generalized complex structure here introduced is identified with the modular matrix $i\hat{\Omega}$ defined in that reference. It can also be compared  with the one  given for the chiral scalars  in the case of  the factorized torus $(T^2)^d$\cite{0701292} and the two coincide when  eq. (\ref{83})  is specified for this  peculiar factorized geometry.

The wave function for the four-dimensional fermions is the solution of the internal Dirac equation:
\begin{eqnarray}
 \gamma^M_{(6)}\tilde{D}_M\eta_n^{ab}=m_n\eta_n^{ab} \,\, ,\label{Dirac}
\end{eqnarray}
being $m_n$ the mass of $n$th-level of the Kaluza-Klein tower  and $\gamma^M_{(6)}$ are the six-dimensional Dirac matrices. In analogy with the dimensional reduction of the bosonic kinetic terms, the eigenfunction problem of the Dirac equation is solved in the complex frame ${\cal Z}$ where both the metric and the magnetic background are diagonal matrices in the non-vanishing off-diagonal blocks. In this complex frame, the Clifford algebra becomes:
\begin{eqnarray}
\left\{\gamma^{{\cal Z}^r},\, \gamma^{\bar{\cal Z}^s}\right\}= 4 \delta^{rs}   \nonumber
\end{eqnarray}
with all the other  anti-commutators vanishing. This algebra is the usual one of fermion creation and annihilation operators and the gamma-matrices can be identified with such operators. According to the identifications (\ref{osc0}), the massless state living in the kernel of the Dirac equation
is obtained by defining a factorized  vacuum  $
\eta_0(\vec{\cal Z}\, , \,{\vec{\bar{\cal Z}}})~=~ u_0~\otimes~\phi_0(\vec{\cal Z}\, ,\,{\vec{\bar{\cal Z}}})$.
Here,  $u_0$ is a constant six-dimensional spinor  and $\phi_0$ is a function of the internal coordinates, both vanishing under the action respectively  of  all the  fermionic and bosonic annihilation operators
\begin{eqnarray}
D_{r}^{({\bar{\cal Z}})}\phi_0({\vec{\cal Z}}, {\vec{\bar{\cal Z}}} )=0~~;~~\gamma^{{\cal Z}^r}_{(6)}u_0=0~~~~{\rm for}~ \lambda_r>0 \nonumber \\
D_{r}^{{({\cal Z})}}\phi_0({\vec{\cal Z}}, {\vec{\bar{\cal Z}}} )=0~~;~~\gamma^{\bar{\cal Z}^r}_{(6)}u_0=0~~~~{\rm for}~ \lambda_r<0 \label{soldirac}
\end{eqnarray}
together with the boundary conditions given in eq. (\ref{bctorus}). The solution of  eq. (\ref{soldirac}) is then obtained  by assuming $\phi_0$ to be  the wave-function in eq. (\ref{83}) and by defining  $u_0= \gamma^{{\cal Z}^r}\chi_0$ for positive eigenvalues $\lambda_r$ and   $u_0= \gamma^{\bar{\cal Z}^r}\chi_0$ for negative eigenvalues,  being $\chi_0$ an arbitrary eight-component constant spinor.

The whole spectrum of the Kaluza-Klein fermions is obtained, according to the standard procedure, by squaring  eq. (\ref{Dirac}):
\begin{eqnarray}
-\left(\gamma^{{\cal Z}^I}_{(6)} D_{I}^{\cal{Z}}\gamma^{{\cal Z}^J}_{(6)} D_{J}^{\cal{Z}}\right)\eta_n & = &
\sum_{r=1}^3\left( |\lambda_r|(2N_r+1) -\frac{1}{4} \left[ \gamma^{{ Z}^r}_{(6)},\,\gamma^{\bar{ Z}^r}_{(6)}\right]|\lambda_r|\right) \eta_n \nonumber \\
&= & (2 \pi R)^{2} m_n^2\eta_n \, , \label{lapl}
\end{eqnarray}
where the bosonic number operator, defined in the previous sections,  has been introduced and the expression of the background gauge field given in eq. (\ref{89}) is used.

The vacuum state shown in eq. (\ref{soldirac}) satisfies the previous equation with $m=0$ and, applying on it an arbitrary number of bosonic oscillators
\begin{eqnarray}
(a_1^\dagger)^{N_1}\,(a_2^\dagger)^{N_2}\,(a_3^\dagger)^{N_3}  \prod_{r=1}^3\gamma^{{Z}^r}\chi_0\otimes \phi_{\vec{j}}(\vec{Z},\,\vec{\bar Z})\,\, ,   \nonumber
\end{eqnarray}
a set of Kaluza-Klein states are generated with masses
\begin{eqnarray}
m^2 =\frac{2}{(2\pi R)^2} \sum_{r=1}^3|\lambda_r| N_r.   \nonumber
\end{eqnarray}
The next levels in the fermion Fock space, satisfying eq. (\ref{lapl}),  are  obtained by applying  one fermion creation operator and an arbitrary number  of  bosonic creation operators:
\begin{eqnarray}
(a_1^\dagger)^{N_1}\,(a_2^\dagger)^{N_2}\,(a_3^\dagger)^{N_3} \gamma^{\bar{Z}^k}  \prod_{r=1}^3\gamma^{{Z}^r}\chi_0\otimes \phi_{\vec{j}}(\vec{Z},\,\vec{\bar Z})~~~~k=1,2,3.  \nonumber
\end{eqnarray}
A tower of KK states is generated with  masses given by:
\begin{eqnarray}
m^2_k=\frac{1}{(2\pi R)^2} \sum_{r=1}^3|\lambda_r|(2N_r)+2\frac{|\lambda_k|}{(2\pi R)^2}~~~~~k=1,2,3 \,\,\,  .   \nonumber
\end{eqnarray}
Other  KK towers  are obtained by acting on the vacuum with two or three fermion creation oscillators and an arbitrary number of bosonic oscillators
\begin{eqnarray}
(a_1^\dagger)^{N_1}\,(a_2^\dagger)^{N_2}\,(a_3^\dagger)^{N_3} \gamma^{\bar{Z}^k} \gamma^{\bar{Z}^l}  \eta_0~~;~~(a_1^\dagger)^{N_1}\,(a_2^\dagger)^{N_2}\,(a_3^\dagger)^{N_3} \prod_{k=1}^3\gamma^{\bar{Z}^k}\eta_0   \nonumber
\end{eqnarray}
with $k,l=1,2,3$. These  are three and one tower of massive states having respectively the same and opposite chirality of the vacuum \cite{1206.3401}. Their mass spectrum is given by:
\begin{eqnarray}
m^2_{k,l}=\frac{2}{(2\pi R)^2} \sum_{r=1}^3|\lambda_r| N_r+2\frac{|\lambda_k|+|\lambda_l|}{(2\pi R)^2}~~;~~m^2=\frac{2}{(2\pi R)^2} \sum_{r=1}^3|\lambda_r| (N_r+1) \,\, .  \nonumber
\end{eqnarray}
All the mass formulas can be collected in a more concise relation by introducing the fermion number operator $N^f_r=0,1$ and by writing
 \begin{eqnarray}
 m^2_n=\frac{2}{(2\pi R)^2} \sum_{r=1}^3|\lambda_r| (N_r+N_r^f) \,\, .\nonumber
\end{eqnarray}
The mass of the Kaluza-Klein fermions coincides  with the one given in eq. (\ref{bms}) valid when the susy condition  $|\lambda_r|+|\lambda_s|=|\lambda_t|$ ($r\neq s\neq t$) is imposed  showing  the consistency and accuracy of the dimensional reduction procedure.

The wave-functions of the chiral matter are  derived  in the background dependent  system of complex coordinates ${\cal Z}_{ab}$
where  the off-diagonal blocks of the background magnetic fields are diagonal.
By definition, in each of these frames a wave-function is associated with the corresponding dy-charged sector of the theory.  The calculation of the effective actions demands the evaluation of overlap integrals among three or more of these functions. It is therefore necessary to re-express such states in terms of quantities defined in a unique system of coordinates as the $w^m$s. In this  frame one has:
\begin{eqnarray}
\eta_0(\vec{w},\,\vec{\bar w})=\prod_{r=1}^3 \left( C^{r}_{s} \frac{ (1+\mbox{sign} \lambda_{r})}{2} \gamma^{w^{s}} + \bar{C}^{r}_{s}
\frac{ (1- \mbox{sign} \lambda_{r})}{2} \gamma^{\bar{w}^{s}}
\right) \chi_0\otimes \phi_{\Omega;\,\vec{j}}(\vec{w},\,\vec{\bar w})\nonumber
\end{eqnarray}
where $C_{r}^{s}, \bar{C}^{r}_{s}$ are the inverse matrices of the ones  defined in eq. (\ref{defcl}) and $\phi_{\Omega;\,\vec{j}}$  is a scalar function of the coordinates. It is defined in eq. (\ref{83}) in terms of the real variables $(x^m,\, y^m)$ and the relation among  these coordinates and the complex ones is given in sect. \ref{2}. By using these relations it is straightforward  to re-write the expression of the wave-function in the complex frame, however the calculus of the Yukawa couplings will be performed in the real system of coordinates and therefore it is not necessary to give a such expression here.
\section{Yukawa Couplings}
\label{4}
The Yukawa couplings are obtained by considering the  trilinear couplings, involving one boson and two fermions, of the ten-dimensional ${\cal N}=1$ SYM action reduced to four dimensions according to the Kaluza-Klein compactification procedure outlined in the previous section. In refs. \cite{0404229,0807.0789,0810.5509}  this dimensional reduction is studied in great detail; here we just quote the result:
\begin{eqnarray}
&&\!\!\!\!S_{3}^{\Phi} =  \frac{1}{2g^2}
\int d^4 x \sqrt{G_{4}}  \int d^6 X^N \sqrt{G_{6}}
\bar{\psi}^{ca}_0({x}^\mu)\,\gamma^5_{(4)} \left[ \varphi_{i,\,0}^{ab}({x}^\mu)\,
\psi_0^{bc}({x}^\mu)\otimes (\eta^{ac}_0)^\dag(x^n,{y}^n)\gamma^i_{(6)}\right.  \nonumber \\
&&\!\!\!\!\times\left.
 \phi^{ab}_{\Omega_{ab};\vec{j}_1}(x^n,{y}^n) \eta_0^{bc}(x^n,{y}^n)  - \varphi_{i,\,m}^{bc}({x}^\mu)\,
\psi_0^{ab }({x}^\mu)\otimes (\eta^{ac}_0)^\dag(x^n,{y}^n)\gamma^i_{(6)} \phi^{bc}_{\Omega_{bc};\vec{j}_2}(x^n,{y}^n) \eta_0^{ab}(x^n,{y}^n)
\right]\nonumber\\
\!\!\!\!\label{S31}
\end{eqnarray}
where $\psi_0$  is the  massless fermion ground state.  $\varphi_{0}$, instead, is the lightest bosonic  excitation which is  massless  if the supersymmetry condition, given soon after eq. (\ref{m2}),  is imposed. In the following, in order to fix notations, we choose $\lambda_1^{ab}$ to be positive. So doing, the massless scalar turns out to be $\phi_{{\cal Z}^1}$, while with the opposite choice $\phi_{\bar{\cal Z}^1}$ would have been the massless state \cite{1206.3401}. In the chosen notations,  only the first term in eq. (\ref{S31}) contributes to the Yukawa coupling for massless particles and one is  left with the expression
\begin{eqnarray}
(S_{3}^{\Phi})^{(1)}= \int d^4 x \sqrt{G_{4}}  \bar{\psi}^{ca}_0\,\gamma^5_{(4)}
\varphi_{{\cal Z}^1,0}^{ab}\, \psi_{0}^{bc} Y^{\vec{j}_1\vec{j}_2\vec{j}_3}
\nonumber
\end{eqnarray}
with the Yukawa coupling constants, in the string frame, given by
\begin{eqnarray}
Y^{ \vec{j}_1\vec{j}_2\vec{j}_3} &=& \frac{1}{2g^2} \left[(u^{ac}_0)^\dag \gamma^{{\cal Z}^1_{ab}}_{(6)} u_0^{bc}\right]{\cal Y}^{\vec{j}_1\vec{j}_2\vec{j}_3}\nonumber
\end{eqnarray}
where
\begin{eqnarray}
{\cal Y}^{\vec{j}_1\vec{j}_2\vec{j}_3}&=&\int_{T^6}d^3x d^3 y \sqrt{G_6}(\phi^{ac}_{\Omega_{ac}; \vec{j}_3}(x^n,\,y^n))^\dag
\phi^{ab}_{\Omega_{ab};\vec{j}_1}(x^n,\,y^n) \phi_{\Omega_{bc};\vec{j}_2}^{bc}(x^n,\,y^n) \,\, .
\label{YY2}
\end{eqnarray}
The integral in eq. (\ref{YY2}) has been computed in ref. \cite{1206.3401}.  The calculation is here summarized in the case in which  all the first Chern-classes associated with the three twisted sectors are independent. When this latter condition is not satisfied there are subtleties that are discussed in ref. \cite{1206.3401}.

The integral can be performed after using the following identity  involving the product of two wave-functions\footnote{In this paper by respect the ref.\cite{1206.3401} a different notation for the indices of the summation is used. The correspondence among the two sets of symbols is: ${\bf Z}_{ab}\equiv  \mathbb{Z}^{3}_{{\rm det}[I_{ab}] I_{ab}^{-1} }$, ${\bf Z}_{bc}\equiv\mathbb{Z}^{3}_{{\rm det}[I_{bc}] I_{bc}^{-1}}$ and $ \tilde{\bf Z}_{ac}= \tilde{\mathbb{Z}}_{(I_{ab}^{-1}+I_{bc}^{-1})\alpha}$}:
\begin{eqnarray}
&&\phi_{\Omega_{ab};\,\vec{j}_1}^{ab}(x^m,\,y^m)~\phi_{\Omega_{bc};\,\vec{j}_2}^{bc}(x^m,\,y^m)={\cal N}_{ab}\,{\cal N}_{bc}\,e^{i\pi \vec{x}^t\,\left(\frac{I_{ab}+I_{bc}}{(2\pi R)^2}\right) \, \vec{ y}+i\pi  \vec{ y}^t \left(\frac{I_{ab}\,\Omega_{ab}+ I_{bc}\,\Omega_{bc}}{(2\pi R)^2}\right) \vec{ y}}\nonumber\\ &&\times\sum_{\begin{array}{l}(\vec{l}_3,\,\vec{l}_4)\in \mathbb{Z}^{3}\\
\vec{m}\in {\tilde{\bf Z}_{ac} }\end{array}} \! \! \! \! \! \sum_{\begin{array}{l}\vec{p}\in{\bf Z}_{bc} \\ \vec{q}\in  {\bf Z}_{ab}  \end{array} }\!\!\!\!e^{i\pi \vec{l}^{~'t} \, Q' \, \vec{l}^{~'}+2\pi i \vec{l}^{~'t} \, Q' \, \left(\begin{array}{c}\vec{y}\\0\end{array}\right)
+ 2\pi i \vec{l}^{~'t} \, {\cal I}' \, \left(\begin{array}{c}\vec{x}\\0\end{array}\right)} \label{phiphi}
\end{eqnarray}
with
\begin{eqnarray}
Q'=
\left(\begin{array}{cc}
                           I_{ab}\Omega_{ab}+I_{bc}\Omega_{bc}& (\Omega_{ab}^t-\Omega_{bc}^t)\alpha^t\\
                           \alpha(\Omega_{ab}-\Omega_{bc})& \alpha(\Omega_{ab}I_{ab}^{-t}+\Omega_{bc}I_{bc}^{-t})\alpha^t\end{array}\right)~~;~~{\cal I}'=\left(\begin{array}{cc}
                              I_{ab}+I_{bc}& (I_{ab}\,I_{ab}^{-t}-I_{bc}\,I_{bc}^{-t})\alpha^t\\
                             0& \alpha( I_{ab}^{-t}+I_{bc}^{-t})\alpha^t
                              \end{array}\right) \,\, ,  \label{q'i'}  \nonumber
\end{eqnarray}
and
\begin{eqnarray}
&&\vec{l}^{~' t}=(\vec{l}_1^{~'t},\, \vec{l}^{~'t}_2)=\left( (\vec{j}_1^{t} I_{ab} +\vec{j}_2^{t}I_{bc} +\vec{m}^t I_{ab})\left(I_{ab} +I_{bc} \right)^{-1}+\vec{l}_3^{\, t};\right.\nonumber\\
&&\left.(\vec{j}_1^t- \vec{j}_2^t+\vec{ m}^t)  \left(I_{ab}^{-1} +I_{bc}^{-1} \right)^{-1} \alpha^{-1}+ \vec{p}^{\, t}\frac{I_{bc}}{{\rm det}[I_{bc}]} +  \vec{q}^{\,t}\frac{I_{ab}}{{\rm det}[I_{ab}]}+   \vec{l}_4^{\, t} \right) . \label{l'}
\end{eqnarray}
The parameter $\alpha$ is chosen  in such a way to make  the matrix  $\alpha(I^{-1}_{ab}+I_{bc}^{-1})$ integer. This request is fundamental in order to obtain the identity written in eq. (\ref{phiphi}). The choice $\alpha={\rm det}[I_{ab}I_{bc}]\mathbb{I}$ satisfies this requirement \cite{0904.0910}.
The indices  of the two summations need some explanation.  Let us  denote by $\mathbb{Z}^3_{(I_{ab}^{-1}+I_{bc}^{-1})\alpha}$ the set of equivalence classes  obtained by  identifying the elements of $\mathbb{Z}^3$ under the shift $\vec{n}\rightarrow \vec{n} +\vec{t}\,(I_{ab}^{-1}+I_{bc}^{-1})\alpha$, with $\vec{t},\vec{n}\in \mathbb{Z}^3$. A subset of $\mathbb{Z}^3_{(I_{ab}^{-1}+I_{bc}^{-1})\alpha}$ is obtained by considering the integer vectors lying
within a cell generated by $\vec{e}_i{\rm det}[I_{ab}]I_{ab}^{-1}$,($i=1,2,3$) being $\vec{e}_i$ defined in the appendix \cite{0904.0910}. This subset is denoted by ${\bf Z}_{ab}$. ${\bf Z}_{bc}$ is defined by exchanging $I_{ab}$ with $I_{bc}$. Finally:
\begin{eqnarray}
\tilde{{\bf Z}}_{ac}=\mathbb{Z}^3_{(I_{ab}^{-1}+I_{bc}^{-1})\alpha}\setminus ( {\bf Z}_{bc } \cup {\bf Z}_{ab }).\label{m} \nonumber
\end{eqnarray}
The proof of the identity written in eq. (\ref{phiphi}) is outlined in the appendix. More details are given in  refs \cite{1206.3401,0904.0910}.

The integral over the $\vec{x}$ variable can be easily performed giving:
\begin{eqnarray}
{\cal Y}^{ \vec{j}_1\vec{j}_2\vec{j}_3} &=& \sqrt{G_6}\,{\cal D}{\cal N}_{ab}\,{\cal N}_{bc}\,{\cal N}_{ca} \sum_{{
\vec{l}_3,\vec{l}_4\in \mathbb{Z}^{3}}}\sum_{\vec{p}\in{\bf Z}_{bc}}\sum_ {\vec{q}\in{\bf Z}_{ab}  } {\cal F}_{\Omega,I}(\vec{l}_3,\vec{l}_4)
e^{i\pi \vec{l}_2^{'t}~ \Pi~ \vec{l}^{~'}_2} \nonumber
\end{eqnarray}
where
\begin{eqnarray}
\Pi=\alpha\left((\Omega_{ab}I_{ab}^{-t}+\Omega_{bc}I_{bc}^{-t})-(\Omega_{ab}-\Omega_{bc})
(I_{ca} \Omega_{ca}+I_{ab} \Omega_{ab}+I_{bc}\Omega_{bc})^{-1}(\Omega_{ab}-\Omega_{bc})^t\right)
\alpha^{-t} \,\,  \label{int}
\end{eqnarray}
and
\begin{eqnarray}
{\cal D}\equiv\sum_{\vec{m}\in\tilde{\mathbb{Z}}_{(I_{ab}^{-1}+I_{bc}^{-1})\alpha}}
\delta_{(\vec{j}_1^t\,I_{ab}+\vec{j}_2^t\,I_{bc}+\vec{m}^t\,I_{ab})(I_{ab}+I_{bc})^{-1}; \vec{j}_3^t}. \nonumber
\end{eqnarray}
The last integral to be computed is contained in the definition of the following function:
\begin{eqnarray}
{\cal F}_{\Omega,I}(\vec{l}_3,\vec{l}_4)\equiv \int_0^1 d^3{{y}} e^{-\pi[\vec{\tilde{y}}^t +\vec{l}^{~'t}_1+\vec{l}^{~'t}_2\, {Q'^{21}}^{t}\,A^{-1}](-iA)[\vec{\tilde{y}}+\vec{l}^{~'}_1+A^{-1} {Q'^{21}}^t\vec{l}^{~'}_2]} \, \nonumber
\end{eqnarray}
with $Q'^{21}=\alpha(\Omega_{ab}-\Omega_{bc})$. The integral is convergent
and, after having  evaluated it, one has the expression \cite{1206.3401}
\begin{eqnarray}
{\cal Y}^{ \vec{j}_1\vec{j}_2\vec{j}_3}  &=&   \int d^3{x} d^3{y}\sqrt{G_6}
{\phi_{\Omega_{ca};\,\vec{j}_1}^{ca}}^*
\phi_{\Omega_{ab};\,\vec{j}_1}^{ab}\, \phi_{\Omega_{bc};\,\vec{j}_2}^{bc}\nonumber\\
&=&{\cal N}_{ab}{\cal N}_{bc}{\cal N}_{ca}\sqrt{G_6}\,{\cal D} \left[{\rm det}(-i(I_{ca}\Omega_{ca}+I_{ab}\Omega_{ab}+I_{bc}\Omega_{bc}))\right]^{-1/2}\nonumber\\
&\times&\sum_{\vec{p}\in {\bf Z}_{bc}}\sum_{\vec{q}\in  {\bf Z}_{ab}}
\Theta\left[\begin{array}{c} \alpha^{-t}I_{bc}^t (\vec{j}_3-\vec{j}_2)+\frac{I_{bc}^t}{{\rm det} I_{bc}^t}\vec{p}+\frac{I_{ab}^t}{{\rm det}I_{bc}}\vec{\tilde{p}}
\\0\end{array}\right](0| \Pi). \nonumber
\end{eqnarray}
that simplifies when all the differences of magnetic fields living on the various stacks of magnetized branes are independent but commuting.  In this approximation, an analogous string calculus of the Yukawa couplings has been performed in ref. \cite{0709.1805}. In this case the quantity $\alpha(I_{ab}^{-1}+I_{bc}^{-1})$ can be made an integer matrix by choosing $\alpha=I_{ab} I_{bc}$ and the product of two wave-functions is still equal to eq. (\ref{phiphi})  specialized  with this value of $\alpha$ and without the sums over the vectors $\vec{p}$ and $\vec{q}$.
The overlap integral over the three wave-functions is now:
\begin{eqnarray*}
{\cal Y}^{ \vec{j}_1\vec{j}_2\vec{j}_3}&=&  \!\!\! \int d^3{x} d^3 {y}\sqrt{G_6}
{\phi_{\Omega_{ca};\,\vec{j}_1}^{ca}}^*
\phi_{\Omega_{ab};\,\vec{j}_1}^{ab}\, \phi_{\Omega_{bc};\,\vec{j}_2}^{bc}
={\cal N}_{ab}{\cal N}_{bc}{\cal N}_{ca}\sqrt{G_6}\, {\cal D} \nonumber\\
&\times&\,\,\left[{\rm det}(-i(I_{ca}\Omega_{ca}+I_{ab}\Omega_{ab}+I_{bc}\Omega_{bc}))\right]^{-1/2}
\,\Theta\!\!\!\left[\begin{array}{c} I_{ab}^{-t}(\vec{j}_3-\vec{j}_2)
\\0\end{array}\right]\!\!(0| {\Pi})
\end{eqnarray*}
with $\Pi$ given by the eq. (\ref{int}) specialized to the value $\alpha=I_{ab}I_{bc}$.

In conclusion, the field theory approach is a very efficient tool in determining  the low-energy effective actions supported in the world-volume of magnetized branes. These coefficients and, in particular, the holomorphic part of the Yukawa couplings strongly depend on the global aspects of the internal manifold as one has explicitly shown in the case of compactifications on  the torus $T^6$. It would be interesting to extend this approach to models  where few global quantities are explicitly computed. In this respect, models coming from compactification of F-theory are a good arena for this kind of analysis.

\appendix
\section{Appendix}
It is useful to give here the proof of eq. (\ref{phiphi}) involving the product of two wave-functions.  According to ref. \cite{MumfordI}
such a product can be written concisely as follows:
\begin{eqnarray}
\phi_{\Omega_{ab};\,\vec{j}_1}^{ab}(x^m,\,y^m )\, \phi_{\Omega_{bc},\,\vec{j}_2}^{bc}( x^m,\,y^m)&=&{\cal N}_{ab}\,{\cal N}_{bc} \,e^{i\pi \vec{ y}^t\,\left(\frac{I_{ab}^t+I_{bc}^t}{(2\pi\,R)^2}\right) \, \vec{ x}+i\pi  \vec{ y}^t \left(\frac{I_{ab}\,\Omega_{ab}+I_{bc}\,\Omega_{bc}}{(2\pi\,R)^2}\right) \vec{ y}}\nonumber\\
&\times&\sum_{\vec{l}\in \mathbb{Z}^{2d}}e^{i\pi \vec{l}^t  Q \, \vec{l}+2\pi i \vec{l}^t  Q  \frac{\vec{ Y}}{2\pi\,R}
+ 2\pi i \vec{l}^t {\cal I}  \frac{\vec{ X}}{2\pi\,R}}   \nonumber
\end{eqnarray}
being $Q={\rm diag}(I_{ab}\Omega_{ab},\,I_{bc}\Omega_{bc})$, ${\cal I}= {\rm diag}(I_{ab},\,I_{bc})$ and
\begin{eqnarray}
\vec{l}=\left(\begin{array}{c}
\vec{n}_1+\vec{j}_1\\
\vec{n}_2+\vec{j}_2\end{array}\right)~~;~~{X}=\left(\begin{array}{c}
                        \vec{ x}\\
                        \vec{ x}\end{array}\right)~~&;&~~{Y}=\left(\begin{array}{c}
                        \vec{ y}\\
                        \vec{ y}\end{array}\right) \,\,\, .    \nonumber
\end{eqnarray}
An equivalent representation of the product of two Riemann Theta functions is obtained by introducing the following transformation matrix \cite{0904.0910}:
\begin{eqnarray}
T=\left(\begin{array}{cc}
        \mathbb{I}&\mathbb{I}\\
        \alpha I_{ab}^{-1}&-\alpha I_{bc}^{-1}\end{array}\right) ~~;~~
        T^{-1}= \left( \begin{array}{cc}
        (I_{ab}^{-1}+I_{bc}^{-1})^{-1}
        I_{bc}^{-1} &(I_{ab}^{-1}+I_{bc}^{-1})^{-1} \alpha^{-1}\\
        (I_{ab}^{-1}+I_{bc}^{-1})^{-1} I_{ab}^{-1} &-(I_{ab}^{-1}+I_{bc}^{-1})^{-1}\alpha^{-1}\end{array}\right) \nonumber
\end{eqnarray}
acting as (see also the relation before  eq.s (\ref{q'i'})):
\begin{eqnarray}
Q'=T  Q  T^t \qquad;\qquad {\cal I}'= T\, {\cal I}\, T^t \,\,\, .\nonumber
\end{eqnarray}
We introduce also the vector:
\begin{eqnarray}
\vec{l}'^{\, t} \equiv \vec{l}^{\, t} T^{-1}&=& \left( (\vec{n}_1+\vec{j}_1)^t(I_{ab}^{-1}+I_{bc}^{-1})^{-1}I_{bc}^{-1} + (\vec{n}_2+\vec{j}_2)^t(I_{ab}^{-1}+I_{bc}^{-1})^{-1}I_{ab}^{-1}\,;\right.\nonumber\\
&& \left.(\vec{n}_1+\vec{j}_1)^t(I_{ab}^{-1}+I_{bc}^{-1})^{-1}\alpha^{-1}
- (\vec{n}_2+\vec{j}_2)^t(I_{ab}^{-1}+I_{bc}^{-1})^{-1}\alpha^{-1}\right)  \,\, .  \nonumber
\end{eqnarray}
By using the following identity:
\begin{eqnarray}
\left( I_{ab}^{-1}+I_{bc}^{-1}\right)^{-1}= I_{bc}\left(I_{ab}+I_{bc}\right)^{-1}I_{ab}= I_{ab}\left(I_{ab}+I_{bc}\right)^{-1}I_{bc}  \nonumber \,\, ,
\end{eqnarray}
one can write \cite{1206.3401}
\begin{eqnarray}
&&\left(\vec{n}_1^t \,I_{ab}+\vec{n}_2^t\,I_{bc}\right)\left(I_{ab}+I_{bc}\right)^{-1}= \vec{m}_1^t \left(I_{ab} +I_{bc} \right)^{-1}+\vec{l}_3^t \nonumber \\
&& \left( \vec{n}_1^t-\vec{n}_2^t \right)I_{ab}\left(I_{ab}+I_{bc}\right)^{-1} I_{bc}\alpha^{-1}= \vec{m}_2^t \left(I_{ab}^{-1}+I_{bc}^{-1}\right)^{-1} \alpha^{-1}+\vec{l}_4^t\label{mdef}
\end{eqnarray}
where $\vec{l}_3,\,\vec{l}_4\in \mathbb{Z}^3$, $\vec{m}_1$ and $\vec{m}_2$ are suitable  integer vectors, while $\alpha$ has to be fixed in  such a way  that the matrix $ \alpha \left(I_{ab}^{-1}+I_{bc}^{-1}\right) $ has integer entries.  In the following, we will choose  $\alpha= {\rm det} \left[ I_{ab}I_{bc}\right] \mathbb{I}$ \cite{0904.0910} which indeed satisfies the above mentioned constraint.
By writing  $\vec{m}_1= m_1^i\vec{e}_i$,  with
\[
\vec{e}_i^t=(\overbrace{0,\dots,0, 1}^{i \,\,\, times},0,\dots) \,\, ,
\]
the lattice with basis vectors $\vec{e}_i \left(I_{ab}+I_{bc}\right)$ is introduced  and, in it,  the equivalent points are those which change $\vec{l}_3$ by   integer values, because this quantity is summed over all the possible elements of $\mathbb{Z}^3$.

$\mathbb Z^{3}_{(I_{ab}+I_{bc})}$ is the set of equivalent classes obtained by identifying the elements of ${\mathbb{Z}^3}$  under the shift $\vec{m}_1+\vec{k}^t\left(I_{ab}+I_{bc}\right)$ ($\forall \vec{k}\in \mathbb{Z}^3$). Inequivalent values of $\vec{m}_1$ lie in the cell determined by the vectors $\vec{e}_i \left(I_{ab}+I_{bc}\right)$ and
their number is $|{\rm det} [I_{ab}+I_{bc}]|$.  Analogously,  the number of inequivalent values of $\vec{m}_2 \in \mathbb Z^{3}_{(I_{ab}^{-1}+I_{bc}^{-1})\alpha}$  is
$|{\rm det }[I_{ab}^{-1}+I_{bc}^{-1}]\alpha|$.

From eqs. (\ref{mdef}),  it is  straightforward to obtain the identities:
\begin{eqnarray}
&&\vec{n}_1^t= (\vec{m}_1^t+\vec{m}_2^tI_{bc})(I_{ab}+I_{bc})^{-1}+\vec{l}_3^{\, t}+\vec{l}_4^{\, t} \alpha I_{ab}^{-1}
\nonumber\\
&&\vec{n}_{2}^t=(\vec{m}_1^t-\vec{m}_2^tI_{ab})(I_{ab}+I_{bc})^{-1} +\vec{l}_3-\vec{l}_4\alpha I_{bc}^{-1} \,\,  \nonumber
\end{eqnarray}
 which are consistent if  both $\alpha I^{-1}_{ab}$ and $\alpha I_{bc}^{-1}$  are  integer matrices. This latter request  is  indeed satisfied by the choice  $\alpha={\rm det}[I_{ab}\,I_{bc}]\mathbb{I}$. Moreover, one has also to impose
\begin{eqnarray}
\vec{m}_1^t+\vec{m}_2^tI_{bc}=\vec{k}^{t} (I_{ab}+I_{bc}) ~~;~~\vec{m}_1^t-\vec{m}_2^tI_{ab}=\vec{k}_{1}^{t}(I_{ab}+I_{bc})   \nonumber
\end{eqnarray}
with $\vec{k}$ and $\vec{k}_1$ elements of $\mathbb{Z}^3$. The solution of the last two equations is
\begin{eqnarray}
\vec{m}_1^t=\vec{m}_2^tI_{ab}+\vec{k}_1^{t}(I_{ab}+I_{bc}) \,  .\label{modi}
\end{eqnarray}
The correspondence between $\vec{m}_1$ and $\vec{m}_2$ is not one-to-one
since the number of the inequivalent values of $\vec{m}_2$  is bigger than the one of inequivalent $\vec{m}_1$. Following ref. \cite{0904.0910}, one can  replace:
\begin{eqnarray}
\vec{m}_{2}^{t}=\vec{\tilde m}_{2}^{t}+\vec{p}^{t} {\rm det}[I_{ab}] (I_{ab}+I_{bc}) I_{ab}^{-1}+ \vec{q}^{t} \, {\rm det}[I_{bc}] (I_{ab}+I_{bc}) I_{bc}^{-1}\nonumber
\end{eqnarray}
and the second line of eq. (\ref{mdef}) becomes:
\begin{eqnarray}
\left( \vec{n}_1^t-\vec{n}_2^t \right)I_{ab}\left(I_{ab}+I_{bc}\right)^{-1} I_{bc}\alpha^{-1} \, \, & =  & \,\,  \vec{\tilde m}_2^t \left(I_{ab}^{-1}+I_{bc}^{-1}\right)^{-1} \alpha^{-1}+\vec{p}^{\,t}\frac{I_{bc}}{{\rm det}I_{bc}} \nonumber
\\ &+ &   \vec{q}^{\, t} \frac{I_{ab}}{{\rm det}I_{ab}}+  \vec{l}_4^{\, t} . \label{newid}
\end{eqnarray}

From eq. (\ref{newid}) one can easily see that shifting $\vec{p} \rightarrow \vec{p} + \vec{k} ({\rm det} [I_{bc}])  [I_{bc}]^{-1}$ for all $\vec{k} \in {\mathbb Z}^{3}$ corresponds to add $\vec{k}$ to $\vec{l}_{4}$, providing equivalent values of $\vec{p}$ since $\vec{l}_{4}$ is summed over all possible integer vectors .  The set of  inequivalent $\vec{p}$ is  denoted by ${\bf Z}_{bc }$ and its number is
$|{\rm det}({\rm det }[I_{bc} ] I_{bc}^{-1})|$. A similar definiton holds for $\vec{q} \in {\bf Z}_{ab}$ and the dimension of this set results to be $|{\rm det}({\rm det }[I_{ab}] I_{ab})^{-1}|$. Consequently, the number of inequivalent $\vec{\tilde m}_2$s is $|{\rm det} [I_{ab}+I_{bc}]|$ which now matches with the one of inequivalent $\vec{m}_{1}$.

By  starting from eq. (\ref{newid}) and repeating the same manipulations which have led to eq. ({\ref{modi}), one has that this latter equation remains unchanged but with $\vec{m}_2$ replaced by  $\vec{\tilde m}_2$. The solution of eq. (\ref{modi})  is now unique and one gets the expression of $\vec{l}'$ given in eq. (\ref{l'}) with $\vec{m}\equiv \vec{\tilde m}_2$.
After collecting all the results,  one derives the identity written in eq. (\ref{phiphi}).

When $I_{ab}$ and $I_{bc}$ commute, the quantities
$\alpha (I_{ab}^{-1}+I_{bc}^{-1})$ can be made an integer matrix with the choice $\alpha= I_{ab}I_{bc}$.  Eqs. ({\ref{mdef}) become:
\begin{eqnarray}
&&\left(\vec{n}_1^t \,I_{ab}+\vec{n}_2^t\,I_{bc}\right)\left(I_{ab}+I_{bc}\right)^{-1}= \vec{m}_1^t \left(I_{ab} +I_{bc} \right)^{-1}+\vec{l}_3^t \nonumber \\
&& \left( \vec{n}_1^t-\vec{n}_2^t \right) \left(I_{ab}+I_{bc}\right)^{-1} = \vec{m}_2^t \left(I_{ab} +I_{bc} \right)^{-1} +\vec{l}_4^t  .  \nonumber
\end{eqnarray}
 with $\vec{m}_{1}, \vec{m}_{2}\in \mathbb{Z}^3_{(I_{ab}+I_{bc})}$. In this case there is no need to introduce the vectors $\vec{p}$ and $\vec{q}$ and one can trivially impose eq. (\ref{modi}).


\begin{thebibliography}{99}

\bibitem{IU} L. E. Ib\`a\~nez and A. Uranga, {\em String Theory and Particle Physics. An Introduction to String Phenomenology,}  Cambridge University  Press (2012).

\bibitem{Review1} R. Blumenhagen, B. K\"ors, D. L\"ust and S. Stieberger, {\em Four-dimensional String Compactifications with D-Branes, Orientifolds and Fluxes}, Phys.Rept. {\bf 445} (2007) 1, [{\tt hep-th/0610327}].
\bibitem{Review2}   M. Cvetic and J. Halverson, {\em TASI Lectures: Particle Physics from Perturbative and Non-perturbative Effects in D-braneworlds}, [{\tt arXiv:1101.2907 [hep-th]}].
%%%%%%%%%%%%%%%%%

\bibitem{magnetized1}
E. S. Fradkin and A. A. Tseytlin, \emph{Nonlinear Electrodynamics from Quantized Strings},
Phys. Lett. {\bf B163} (1985) 123.
\bibitem{magnetized3}
A. Abouelsaood, C. G. Callan Jr., C. R. Nappi and S. A. Yost,\emph{ Open Strings in Background Gauge Fields}, Nucl. Phys. {\bf B280} (1987) 599.
\bibitem{BachasPorrati} C. Bachas and M. Porrati, {\em Pair creation of open strings in an electric fields}, Phys. Lett. {\bf B296} (1992) 77, [{\tt hep-th/9909032}];
\bibitem{0512067}
M. Bertolini, M. Bill\`{o}, A. Lerda, J. F. Morales and Rodolfo Russo, \emph{Brane world effective actions for D-branes with fluxes},
Nucl. Phys. {\bf B743} (2006) 1, [{\tt hep-th/0512067}].
  \bibitem{0709.4149}
P. Di Vecchia, A. Liccardo, R. Marotta, I. Pesando and  F. Pezzella, \emph{Wrapped Magnetized Branes: Two Alternative Descriptions?}, JHEP {\bf 0711} (2007) 100,
[{\tt arXiv:0709.4149 [hep-th]}].
\bibitem{also6}
    H. Abe, T. Kobayashi, H. Ohki and K. Sumita, \emph{     	
Superfield description of 10D SYM theory with magnetized extra dimensions}, [{\tt arXiv:1204.5327 [hep-th]}].
%%%%%%%%%%%%%%%%%%
\bibitem{0508043}
M. Berg, M. Haack and B. Kors, \emph{String Loop Corrections to K\"ahler Potentials in Orientifolds}, JHEP {\bf 0511} (2005) 030	[{\tt hep-th/0508043}].
\bibitem{also4}
P. Di Vecchia, A. Liccardo, R. Marotta and F. Pezzella, {\em K\"ahler Metrics: String vs Field Theoretical Approach},    Fortsch.Phys. {\bf 57} (2009) 718,  [{\tt arXiv:0901.4458v1 [hep-th]]}.
\bibitem{also5}
 H. Abe, K. S. Choi, T. Kobayashi and  H. Ohki, \emph{
    Higher Order Couplings in Magnetized Brane Models
    Hiroyuki Abe, Kang-Sin Choi, Tatsuo Kobayashi, Hiroshi Ohki} JHEP {\bf 0906} (2009) 080,
    [{\tt  arXiv:0903.3800  [hep-th]}].

%%%%%%%%%%%%%%%%%%%%%%%%%%%%%%%%%%%%%%%%%%%%%%



%%%%%%%%%%%%%%%%%%%%%%%%
\bibitem{0404229}
D. Cremades, L. E. Ibanez and F. Marchesano, \emph{Computing Yukawa
couplings from magnetized extra dimensions},
JHEP {\bf 0405} (2004) 079,
[{\tt hep-th/0404229}].
\bibitem{0701292}
R. Russo and S. Sciuto,
\emph{The twisted open string partition function and Yukawa couplings},
 JHEP {\bf 0704} (2007) 030,  [{\tt hep-th/0701292}].
\bibitem{0807.0789}
J. P. Conlon, A. Maharana and F. Quevedo, \emph{Wave Functions and Yukawa Couplings in Local String Compactifications}, JHEP {\bf 0809} (2008) 104,
[{\tt arXiv:0807.0789  [hep-th]}].
\bibitem{0810.5509} P. Di Vecchia, A. Liccardo, R. Marotta and F. Pezzella, \emph{K\"{a}hler Metrics and Yukawa Couplings in Magnetized Brane Models}, JHEP {\bf 0903} (2009) 029,
[{\tt arXiv:0810.5509  [hep-th]}].
\bibitem{0906.3033}
 P. G. C\'amara and F. Marchesano, {\em Open string wavefunction in flux compactification},  JHEP 0910 (2009) 017,  [{\tt arXiv:0906.3033 [hep-th]}].

%%%%%%%%%%%%%%%%%%%%%%%%
\bibitem{1206.3401} L. De Angelis, R. Marotta, F. Pezzella, R. Troise, {\em More About Branes on a General Magnetized Torus}, JHEP 1210 (2012) 052, [{\tt  	arXiv:1206.3401}].


\bibitem{1101.0120}P. Di Vecchia, R. Marotta, I. Pesando and F. Pezzella, {\em Open strings in the system D5/D9}, J. Phys.  {\bf A44} (2011), 245401, [{\tt arXiv:1101.0120 [hep-th]}].

\bibitem{0904.0910}
I. Antoniadis, A. Kumar and B. Panda, { \em Fermion Wavefunctions in Magnetized branes: Theta identities and Yukawa couplings}, Nucl. Phys. {\bf B823} (2009) 116, [{\tt arXiv:0904.0910 [hep-th]}].



\bibitem{MumfordI} D. Mumford, {\em Tata Lectures
on Theta I}, Birkh\"auser, Boston 1983.

\bibitem{0709.1805} D. Duo, R. Russo and S. Sciuto, {\em  New twist field couplings from the partition function for multiply wrapped D-branes}, JHEP {\bf 0712} (2007) 042, [{\tt arXiv:0709.1805 [hep-th]}].

\bibitem{THooft} G. W. 't Hooft, {\em Some twisted self-dual solutions for the Yang-Mills equations
on a hypertorus}, Commun. Math. Phys. {\bf 81} (1981) 267.	
\bibitem{0306006} M. Sakamoto and S. Tanimura, {\em An Extension of Fourier analysis for the n torus in the magnetic field and its application to spectral analysis of the magnetic Laplacian},  J. Math. Phys. {\bf 44} (2003) 5042, [{\tt hep-th/0306006}].


























\end{thebibliography}
\end{document}